\documentclass{optica-article}
\DeclareMathOperator*{\argmin}{arg\,min}
\DeclareMathOperator*{\argmax}{arg\,max}

\journal{opticajournal} % for journals or Optica Open

\articletype{Research Article}

\usepackage{lineno}
\usepackage{pdfpages}
\begin{document}
\title{Definition of Multi-Parameter Super-Resolution for Optical Telescopes}

\author{Peter Menart\authormark{1,$\dag$},
Seungman Choi\authormark{1,$\dag$} and Zubin Jacob\authormark{1, *}}

\address{
\authormark{1}Elmore Family School of Electrical and Computer Engineering, Birck Nanotechnology Center, Purdue University, West Lafayette, Indiana 47907, USA\\
\authormark{$\dag$}These authors contributed equally to this work.\\
\authormark{*}zjacob@purdue.edu\\
}

\begin{abstract*}
Super-resolution techniques in microscopy are opening new frontiers in biology, and super-resolution for telescope imaging may provide similar benefits for astronomy and space domain awareness. Telescope super-resolution techniques such as multi-plane light conversion and photonic lanterns have been inspired by the ability of mode sorting measurements to extract classical Fisher information closer to the quantum bound than standard telescope imaging. Fisher information analyses for super-resolution imaging systems commonly consider the case of imaging point sources, but often only consider estimating a small subset of the parameters. Furthermore, there has been no demonstration of an estimation algorithm with performance that follows the limit set by the Fisher information when all point source parameters are considered. We provide a framework for defining and analyzing super-resolution using the classical and quantum Fisher information for the multi-parameter estimation problem and demonstrate an algorithm that can achieve the classical Fisher information bound for all the parameters simultaneously. 
\end{abstract*}

\section{Introduction} Super-resolution refers to imaging objects separated by a distance smaller than the resolution limit of the imaging system. In microscopy, super-resolution imaging is mature and is driving scientific discoveries \cite{benard2024combining, SRMreview, STED}. Super-resolution for optical telescopes requires fundamentally different techniques which are still under development. Notably, telescopes are used for imaging targets that are far away, making active illumination of the target, a standard strategy in microscopy, impossible. As a result, the light field to be measured is defined entirely by the object being imaged, and telescope super-resolution techniques often focus on extracting as much information from this light field as possible. Notably, the information perspective indicates the possibility of imaging objects with separation smaller than the Rayleigh limit, which is the traditional way to quantify imaging system resolution \cite{Rayleigh}.

\begin{figure}[ht]
\centering
\includegraphics[width=0.7\textwidth]{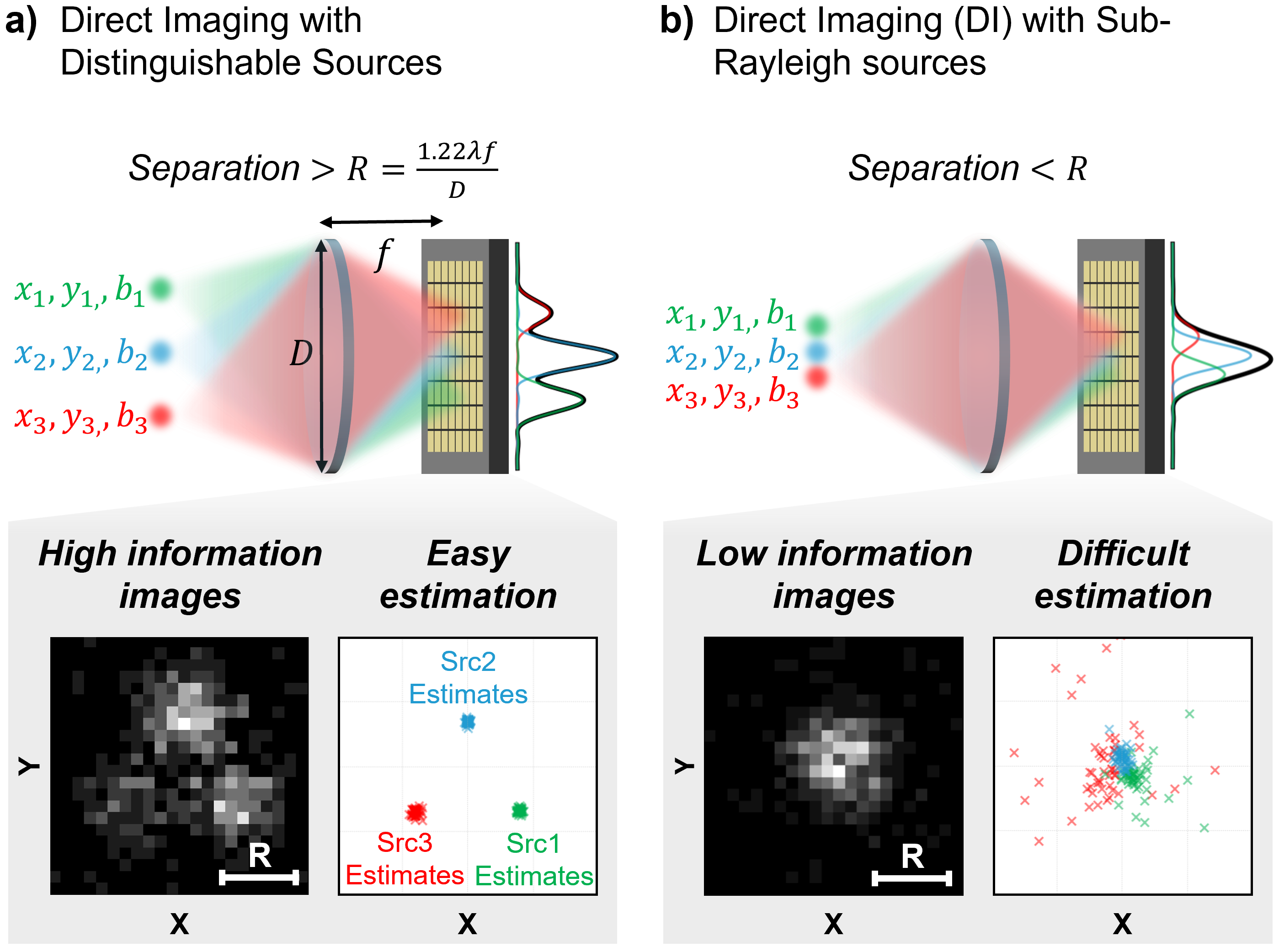}
\caption{We consider the problem of estimating the positions $(x_n, y_n)$ and brightnesses $b_n$ of multiple point sources with a direct imaging (DI) measurement. (a) When the separation of the sources is larger than the Rayleigh limit (R), the sources are easily distinguishable and estimating these parameters is relatively easy. (b) When the separation is less than the Rayleigh limit, the overlap of the point sources reduces the information in the images and estimation is challenging.}
\label{fig:Fig1}
\end{figure}

\begin{figure*}[ht]
\centering
\includegraphics[width=\textwidth]{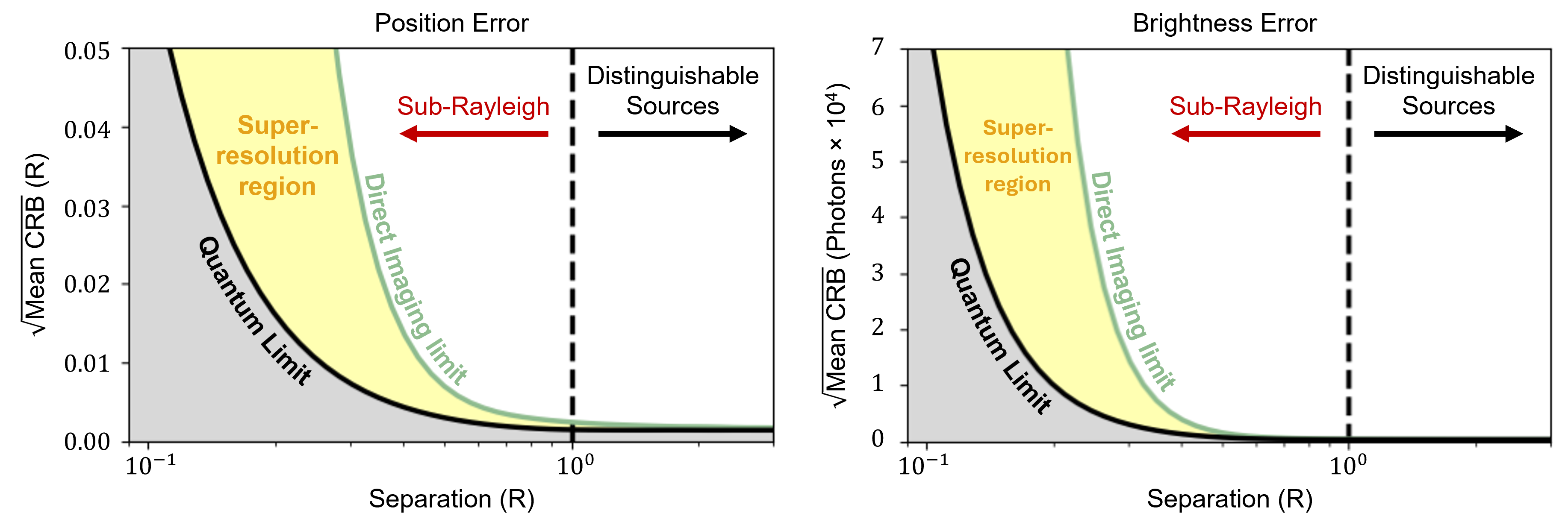}
\caption{The fundamental information bound for imaging five point sources. Each curve shows the bound on the collective RMSE of the position/brightness parameters for all five sources. Super-resolution requires a measurement with a lower bound than DI. The quantum limit shows the minimum error possible for any imaging system. Note that as the separation goes below the Rayleigh limit, both the DI error and the quantum limited error start to diverge, reflecting that the sources become indistinguishable at zero separation. The sources are arranged in an 'X' shape with relative brightnesses [1, 2, 3, 4, 5] and total photon number 5e5. The separation on the X axis refers to the minimum separation between sources.}
\label{fig:Fig2}
\end{figure*}

To quantify the amount of information in a measurement, we must recognize that imaging is a multi-parameter estimation problem. In the super-resolution literature, the case of imaging $N$ point sources is commonly considered. The parameters to be estimated are then the positions and brightnesses of the point sources, as shown in Fig. \ref{fig:Fig1}. The information extracted by an imaging system can be quantified by the classical Fisher information, whose inverse determines the minimum error with which the parameters can be estimated from the given measurement. More information corresponds to a lower minimum error. Similarly, the information contained in the light that enters the telescope is described by the quantum Fisher information, which is the maximum amount of classical Fisher information that any measurement can extract. 

Analyses of super-resolution imaging systems such as multi-plane light conversion, photonic lanterns, and image inversion interferometers have focused on estimating the separation of two point sources or some subset of the point source parameters \cite{tsang2016, ImgInvFluor, BoydAxial, ExpCoronGuhaAshok, paur2018tempering, larson2021two, HomodyneFILvovsky, GuhaArbPSF, LLSoto2018optimal, Prasad2018quantum, santamaria2024single}. Similarly, quantum Fisher information calculations have often focused on a parameter subset, been restricted to just two point sources, or assumed one-dimensional imaging \cite{tsang2016, Prasad2018quantum, AshokQuantLimExo, vrehavcek2017multiparameter, liang2023quantum}. 
% Classical and quantum Fisher information analyses in the super-resolution literature have often focused on estimating the separation of two point sources, or some subset of the point source parameters \cite{tsang2016, paur2018tempering, ExpCoronGuhaAshok, AdModeSortZipp, santamaria2024single, liang2023quantum, vrehavcek2017multiparameter}. 
We note that from the Fisher information perspective, excluding parameters is equivalent to assuming the parameter values are already known with no uncertainty. Additionally, since the estimation of each parameter is generally coupled to the estimation of the other parameters, excluding any parameter affects the minimum errors for the remaining parameters. Thus, it is more realistic to include all of the point source parameters in the Fisher information analysis. 
Furthermore, while the classical Fisher information limit has been used to evaluate imaging systems, there has been no demonstration that this classical limit is achievable when estimating all of the point source parameters.

In this paper, we provide a framework for analyzing super-resolution systems for the case of point source imaging. Our framework is based on classical and quantum Fisher information calculations that are applicable for arbitrary configurations of any number of point sources, and which assume that all point source positions and brightnesses must be estimated. We observe that the traditional Rayleigh limit is not fundamental, as standard telescope imaging, also called direct imaging (see Fig. \ref{fig:Fig1}), can image sources separated by less than the Rayleigh limit \cite{choi2024telescope}. Consequently, we argue that super-resolution is better defined by achieving error less than the Fisher information limit for direct imaging (DI). This requires an imaging system that extracts more Fisher information from the incoming light than DI, along with an estimation algorithm that can utilize this additional information. Additionally, we demonstrate that a natural gradient descent estimation algorithm can achieve the classical Fisher information limit for all parameters simultaneously for a DI system. This algorithm can be easily extended to super-resolution imaging systems using automatic differentiation. The code for the Fisher information calculations and the natural gradient descent algorithm is available on Github (Ref. \cite{CodeRepo}).

% Additionally, we demonstrate that the classical Fisher information limit for DI can be achieved simultaneously for all parameters using a natural gradient descent algorithm, justifying the use of this DI limit as a benchmark. This also provides a possible approach to extract information from super-resolution measurements that 

\section{Definition of Multi-Parameter Super-Resolution}
To formulate the Fisher information criteria for super-resolution, we view imaging as a parameter estimation problem as shown in Fig. \ref{fig:Fig1}. We consider imaging $N$ point source objects, each with position $(x_n, y_n)$ and brightness $b_n$. We thus want to estimate the parameter vector $\boldsymbol{\theta} = (x_1, y_1, b_1, ..., b_N)$, which has $3N$ parameters. We first consider the case of a direct imaging (DI) measurement. Note that when the separation between the sources is larger than the Rayleigh limit, the sources have little overlap and estimating these parameters is relatively straightforward, as shown in Fig. \ref{fig:Fig1}a. However, when the sources are much closer than the Rayleigh limit, the images of the point sources overlap on the detector, resulting in images with reduced information, and estimating the $3N$ parameters is more difficult. See Fig. \ref{fig:Fig1}b. We propose defining super-resolution imaging as an imaging technique that can estimate the source parameters more accurately than is possible with DI. This is illustrated in Fig. \ref{fig:Fig2}, where super-resolution corresponds to estimating the positions and/or brightnesses with less error than the DI limit. 

We will now explain the calculation of this limit. The Fisher information matrix $\mathbf{J}$ is given by

\begin{equation}
J_{lk}(\boldsymbol{\theta}) = \mathbb{E}\left[ \frac{\partial \ln(p(\mathbf{I}; \boldsymbol{\theta}))}{\partial \theta_l} \frac{\partial \ln(p(\mathbf{I}; \boldsymbol{\theta}))}{\partial \theta_k} \; \middle\vert \; \boldsymbol{\theta}\right],
\label{eq:FIM}
\end{equation}

where $\mathbb{E}\left[\cdot \; \middle\vert \; \boldsymbol{\theta}\right]$ denotes the expected value given the parameters $\boldsymbol{\theta}$, $\mathbf{I}$ is the vector of measurements, $p(\mathbf{I}; \boldsymbol{\theta})$ is the probability distribution of the measurement given $\boldsymbol{\theta}$, and the derivatives are evaluated at the true parameter values $\boldsymbol{\theta}$ \cite{FItutorial}. Each element of $\mathbf{I}$ is the number of photons measured at the corresponding detector pixel. The Cram\'er-Rao lower bound (CRB) relates the Fisher information matrix to the minimum variance of any unbiased estimator of the parameters as

\begin{equation}
\mathrm{Var}\left(\widehat{\theta_i}\right) \geq \left[\mathbf{J(\boldsymbol{\theta})}^{-1}\right]_{ii} = \mathrm{CRB}(\theta_i),
\label{eq:CRB}
\end{equation}
where $\widehat{\theta_i}$ denotes an estimate of parameter $\theta_i$ \cite{Kay}. Note that since the estimator is assumed to be unbiased, the mean square error is equal to the variance, and the square root of the CRB provides a lower bound on the root mean square error (RMSE) of the parameter estimate. 

\begin{figure*}[t]
\centering
\includegraphics[width=\textwidth]{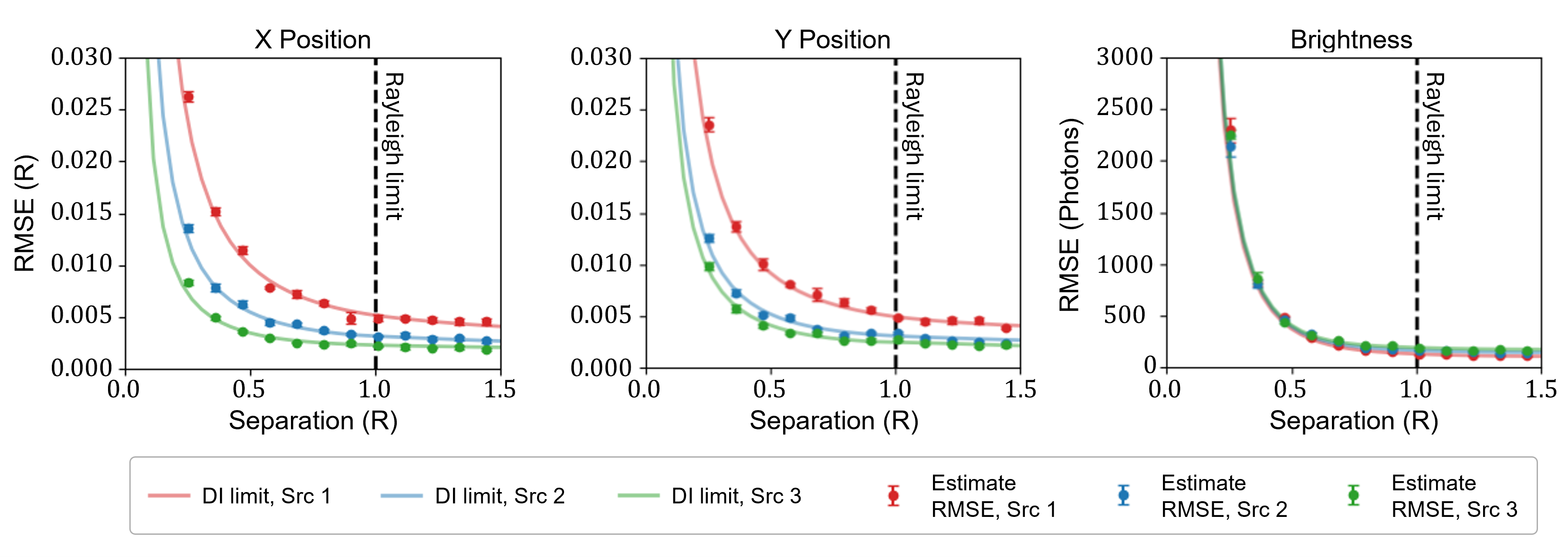}
\caption{Demonstration of the achievability of the DI limit for a three source case. The estimates are produced using a maximum likelihood estimator on simulated images, where we assume the measurement is shot-noise limited. The plots show data from 2000 trials with different separations, with the data points showing the RMSEs of the estimates in a given separation bin. The RMSEs of the estimates follow the DI limit, indicating the maximum likelihood estimator has minimal error for these scenes. Error bars are the standard deviation of the RMSE calculated by splitting the trials into batches. The relative brightness of the sources is [1, 2, 3] with 5e4 total photons in the images. The sources are arranged in an equilateral triangle.}
\label{fig:Fig3}
\end{figure*}

Note the diagonal component $J_{ii}$ specifies the sensitivity of the measurement to the parameter $\theta_i$, and a larger $J_{ii}$ corresponds to a smaller $\mathrm{CRB}(\theta_i)$. The off diagonal components $J_{ki}$ with $k\neq i$ describe the effect of the uncertainty in the $\theta_k$ estimate on the $\theta_i$ estimate. Non-zero off-diagonal elements always increase $\mathrm{CRB}(\theta_i)$, and having more unknown parameters will always give an equal or larger $\mathrm{CRB}(\theta_i)$ \cite{Kay}. Thus, the CRB for each parameter depends on the entire estimation problem.

% Importantly, there are two contributions to the CRB for each parameter $\theta_i$. The diagonal component $J_{ii}$ specifies the sensitivity of the measurement to the parameter $\theta_i$. A larger $J_{ii}$ corresponds to a smaller $\mathrm{CRB}(\theta_i)$. The off diagonal components $J_{ik}$, $J_{ki}$ with $k\neq i$ describe the affect of the uncertainty in the $\theta_k$ estimate on the $\theta_i$ estimate. Non-zero off-diagonal elements always increase $\mathrm{CRB}(\theta_i)$, and having more unknown parameters will always give an equal or larger $\mathrm{CRB}(\theta_i)$ \cite{Kay}. Thus, the CRB for each parameter depends on the entire estimation problem.

To calculate the CRB, we utilize the standard assumption for telescope imaging of weak thermal sources, which results in the photon numbers at the detector pixels following independent Poisson distributions, often referred to as shot noise. We ignore detector noise as optical detectors are now approaching shot noise limited performance and detector noise is not a fundamental limitation. The Fisher information matrix is then given by

\begin{equation}
J_{lk}(\boldsymbol{\theta}) = \sum_{p}{\frac{\partial \mu_p(\boldsymbol{\theta})}{\partial \theta_l} \frac{\partial \mu_p(\boldsymbol{\theta})}{\partial \theta_k} \frac{1}{\mu_p(\boldsymbol{\theta})}},
\label{eq:FIMPoiss}
\end{equation}

where the sum is over all pixels and $\mu_p(\boldsymbol{\theta})$ is the mean photon number at each pixel $p$, which is dictated by the imaging system \cite{motka2016optical, FisherMicroTutorial}. We will denote the mean photon number at each pixel for direct imaging as $\mu_p^{DI}$. 

Using Eq. \ref{eq:FIMPoiss} for the FIM and the DI measurement model, we can calculate the CRB for direct imaging, $\mathrm{CRB}^{DI}$. See the supplementary material for details. The quantity $\sqrt{\mathrm{CRB}^{DI}}$ is then the lower bound on the RMSE for any unbiased estimator using data from a DI measurement. We define a super-resolution technique as a measurement system, described by $\mu^{SR}$, that has a lower CRB than DI, paired with an estimator $\widehat{\boldsymbol{\theta}}^{SR}$ that achieves an RMSE lower than the DI limit,

\begin{equation}
 \sqrt{\mathrm{CRB}^{SR}(\theta_i)} \leq \mathrm{RMSE} \left(\widehat{\theta_i^{SR}}\right) < \sqrt{\mathrm{CRB}^{DI}(\theta_i)} .
\label{eq:SuperRes}
\end{equation}

We assume that the CRBs are calculated using the same telescope aperture size and integration time for a fair comparison. Any initial measurements used to inform the super-resolution technique should also be included when equalizing the integration time \cite{QAI, AdModeSortZipp}. Examples of measurement systems that could be used instead of DI include multi-plane light conversion devices, photonic lanterns, and interferometric measurements \cite{eikenberry2024photonic, OnSkyPL, larsonInvInt, ExpCoronGuhaAshok}. Note that the super-resolution estimator RMSE may violate the DI bound for some or all of the parameters, but the estimator RMSE should not violate the super-resolution measurement bound for any parameter. This would require the estimator to be biased. In general, it is possible to obtain error below the CRB using a biased estimator, but this does not necessarily correspond to extracting meaningful information from the measurements. Biased estimation is discussed more below. Fig. \ref{fig:Fig2} shows the direct imaging bounds for the position and brightness RMSE when imaging 5 point sources with unequal brightnesses. Note that the DI limits diverge as the separation goes to zero, reflecting the fact that the sources become indistinguishable as the separation becomes small.

% Importantly, we have included all the point source positions and brightnesses in our CRB calculation. In some super-resolution techniques, initial measurements are made which are then used to inform the super-resolution technique \cite{QAI, AdModeSortZipp}. Rather than take parameters estimated from previous measurements to be known, these initial measurements should simply be included in the Fisher information calculation and considered when equalizing the integration time.

It is further possible to bound the maximum degree of super-resolution achievable using the quantum Fisher information matrix $\boldsymbol{\mathcal{F}}(\boldsymbol{\theta})$, which is calculated as 

\begin{equation}
\mathcal{F}_{lk}(\rho) = \mathrm{Re}[\mathrm{Tr}( \mathcal{L}_l\mathcal{L}_k \rho)] ,
\label{eq:QFIM_SLD}
\end{equation}

where $\rho$ is the density operator of the light entering the telescope and $\mathcal{L}_k$ is the symmetric logarithmic derivative operator for parameter $\theta_k$ \cite{tsang2016}. The inverse of $\boldsymbol{\mathcal{F}}(\boldsymbol{\theta})$ gives the quantum Cram\'er-Rao lower bound (QCRB), which bounds the classical CRB and estimator variance as 

\begin{equation}
\mathrm{Var}\left(\widehat{\theta_i}\right) \geq \left[\mathbf{J(\boldsymbol{\theta})}^{-1}\right]_{ii} \geq \left[\boldsymbol{\mathcal{F}}(\boldsymbol{\theta})^{-1}\right]_{ii}
\label{eq:QCRB}
\end{equation}

We calculate $\boldsymbol{\mathcal{F}}(\boldsymbol{\theta})$ numerically as described in Appendix A. We note that we include all the point source positions and brightnesses and the calculation is general for any number of point sources, which differs from previous point source QFI calculations.

The QCRB thus gives the minimum uncertainty over all possible measurement systems $\mu$. Fig. \ref{fig:Fig2} shows the quantum limit for estimator RMSE compared to the DI RMSE limit. Notably, the ratio of the DI limit to the quantum limit increases as the separation decreases, indicating that super-resolution techniques are primarily useful for small separations. In this case, at 0.25R separation, the quantum limit is more than 6x smaller than the DI limit, indicating significant room for improvement by super-resolution techniques. This ratio decreases as the separation gets larger, and there is not much room for improvement when the separation is large.

% We note that for small separations, the quantum limit is significantly below the DI limit, indicating that super-resolution measurements can have a significant advantage. As the source separation gets larger, the ratio of the DI CRB to the QCRB gets closer to one and there is not much room for improvement.

\section{Achieving the Multi-Parameter CRB} 

\begin{figure}[t]
\centering
\includegraphics[width=0.7\linewidth]{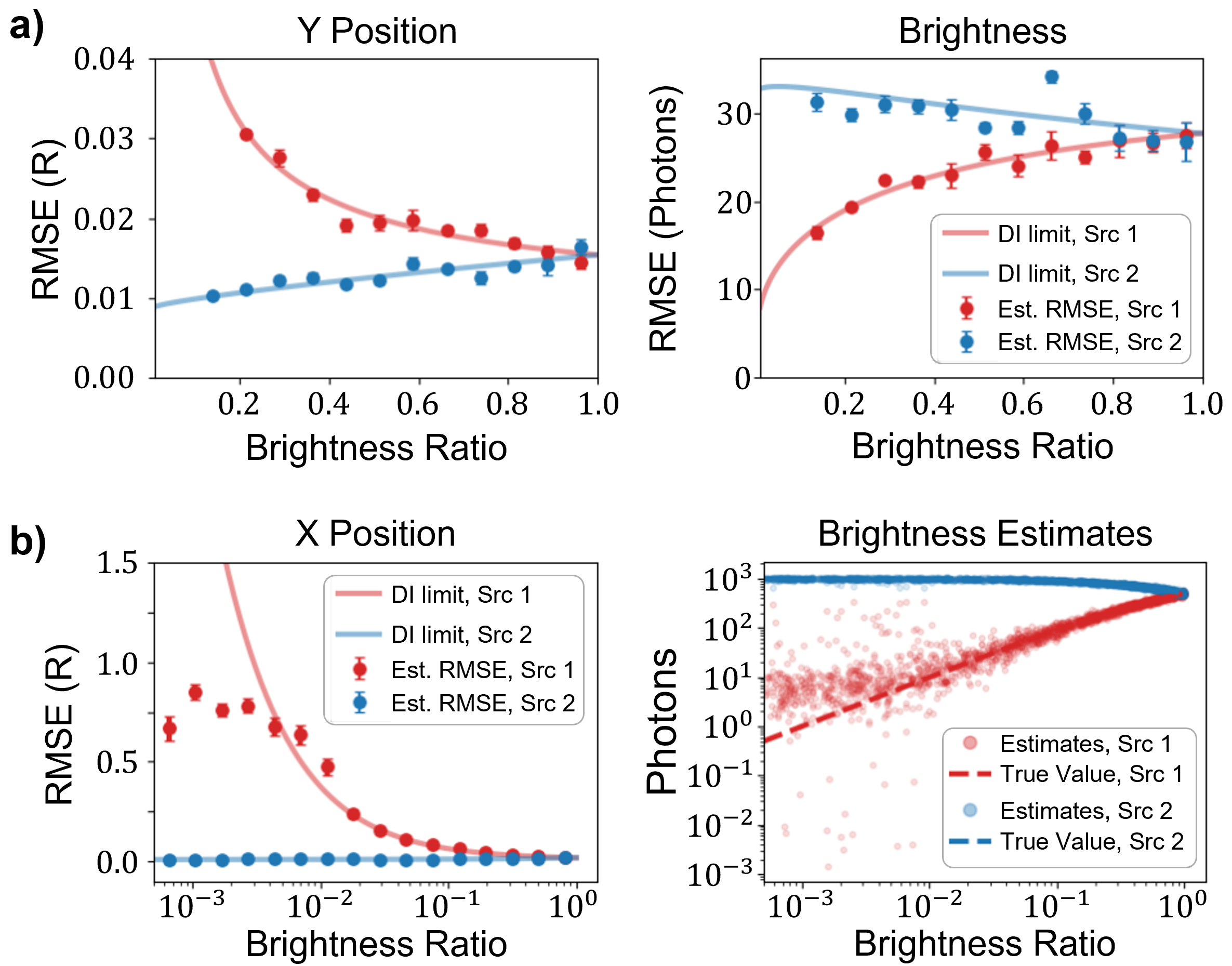}
\caption{Contrast resolution for a 2 point source scene with a DI measurement. The sources are 0.8R apart and the mean total photon number is 1000. Each plot shows RMSEs obtained from 2500 Monte Carlo trials. (a) The Y position and brightness errors for the ML estimates match the theoretical limit for this scenario, again demonstrating the achievability of the DI limit. The minimum brightness ratio for the Monte Carlo trials is 0.1. (b) The X position errors also follow the DI limit down to ~0.02 contrast, in accordance with the top panel. At higher contrast, the DI limit is violated because the MLE estimate fits to noise and becomes biased, as indicated by the brightness estimates.}
\label{fig:Fig4}
\end{figure}

Mathematically, the maximum likelihood estimator (MLE) is known to achieve the CRB in the asymptotic limit of high signal to noise ratio \cite{Kay}. However, it is generally not known how much signal is required to achieve the CRB, and finding the MLE often involves a challenging optimization problem. For Poisson noise, the MLE is given by

\begin{equation}
\widehat{\boldsymbol{\theta}^{ML}} = \argmax_{\boldsymbol{\theta}}{p(\mathbf{I}; \boldsymbol{\theta})} = \argmin_{\boldsymbol{\theta}} \sum_{p} \mu_p(\boldsymbol{\theta}) -I_p \ln( \mu_p(\boldsymbol{\theta})) . 
\label{eq:MLE}
\end{equation}

We solve this optimization problem using a natural gradient descent algorithm \cite{NGref}. Gradient descent algorithms are prone to converging to local extrema, so we are not guaranteed to find the global maximum. This can result in failure to achieve the CRB and can cause bias. To increase our chance of finding the true MLE, we randomly initialize the algorithm multiple times and take the estimate with the largest likelihood as the final estimate. See Appendix B for details on the algorithm.

Fig. \ref{fig:Fig3} shows the direct imaging limit compared to the RMSE of the MLE estimates obtained from the natural gradient algorithm for 2000 simulated direct imaging measurements. We note that the RMSEs closely match the direct imaging limit for all parameters, demonstrating the achievability of the DI CRB using this approach. Since the DI limit is attainable for all parameters simultaneously, it is reasonable to compare super-resolution methods directly to the DI limit.

Fig. \ref{fig:Fig4}a gives another example of the achievability of the DI CRB, and illustrates the utility of the Fisher information for quantifying resolution in parameters other than the separation. Fig. \ref{fig:Fig4}a shows the DI limit and MLE RMSEs for different brightness ratios of two point sources. Small brightness ratios correspond to a high contrast scene. Similar to the separation case, the CRBs for the dim source diverge as the brightness ratio goes to zero, indicating that at some point the dim source is not detectable. Naturally, photon number resolution could also be analyzed by considering different total photon numbers. 

In Fig. \ref{fig:Fig4}b, we plot the RMSEs of the MLE for the X position over a large range of contrasts, and we observe that at brightness ratios lower than 0.01, the RMSEs violate the DI limit. This indicates that the MLE must become biased in this region. The estimate bias is clear in the brightness estimates, which do not follow the true values for the dim point source at low brightness ratios. While this results in error lower than the theoretical limit, it does not necessarily mean these estimates are useful. The flattening out of the brightness estimates reflects that for these brightness ratios, the dim point source is simply fit to noise from the bright point source, and the estimates do not correlate to the true parameters describing the dim source. Estimation bias can also occur intentionally when prior information is used in the estimation algorithm. In this paper, we focus on unbiased estimation to emphasize super-resolution techniques as methods that fundamentally extract more information from the incoming light than direct imaging.

% rather than comparing performance over many trials.

% This demonstrates that it is possible to achieve sub-Rayleigh imaging performance with a normal direct imaging measurement and gradient descent post-processing, making it clear that merely imaging sources below the Rayleigh limit should not constitute super-resolution
\section{Discussion and Conclusion}

In this paper, we have considered the scenario of imaging point sources. This simplifies calculations and facilitates a comparison with the Rayleigh limit, which was defined in the context of imaging two point sources. However, the Fisher information calculation can be extended to more general imaging scenarios such as extended object imaging \cite{gong2026passive, AdModeSortZipp}. Further generalizations of the framework could also incorporate the need to estimate the effect of atmospheric turbulence, which causes distortion in images from ground-based telescopes. Atmospheric turbulence parameters can also be incorporated into the estimation problem and the Fisher information framework \cite{TAPBD, PWI}. 

% A number of works have also considered the distance of the point sources from the telescope as a parameter of interest \cite{BoydAxial, Prasad2018quantum}. In astronomy, the targets are so far away that the distance is unlikely to affect the measurements or be able to be estimated, but in other applications the target distance may be relevant.

Finally, we note that to conduct the Fisher information analysis, we have assumed the number of point sources in the scene is known. Determining the number of sources in the scene is a detection problem, and Fisher information analysis cannot be applied in the same manner as for the estimation of continuous parameters. Utilizing detection theory to include the source number estimation in the theoretical framework would be desirable. Alternatively, using an extended source parameterization as discussed above would also resolve this issue.

In conclusion, we have presented a Fisher information framework for analyzing super-resolution for point source imaging where we do not assume any positions or brightnesses are known, and have shown one method to achieve the fundamental error limit for DI for all parameters simultaneously.

% In conclusion, we have presented a definition for super-resolution, which requires both a measurement system that extracts more Fisher information than direct imaging and an estimator that utilizes that information. 

\appendix

\section*{Appendix A: Numerical Calculation of the Quantum Fisher Information}

In this appendix, we explain our numerical approach for calculating the QFI in Eq. \ref{eq:QFIM_SLD}. We use the following expression for the density operator

\begin{equation}
\rho = \sum_{n=1}^{N} b_n |\psi_n\rangle \langle\psi_n|,
\label{eq:rho_final}
\end{equation}

where $|\psi_n\rangle$ is the quantum state representing the electric field from source $n$. See the supplementary material for a derivation of this density operator. The SLD operator in Eq. \ref{eq:QFIM_SLD} is defined by the equation $\partial_k \rho = \frac{1}{2}(\rho \mathcal{L}_k + \mathcal{L}_k \rho)$, and we calculate it using the expression 

\begin{equation}
\mathcal{L}_{k}(\rho) = \sum_{i, j = 0, E_i + E_j \neq 0}^{d-1} \frac{2\mathrm{Re}(\langle e_i | \partial_k \rho |e_j \rangle)}{E_i + E_j} ,
\label{eq:SLD}
\end{equation}

where the $E_i, |e_i\rangle$ are the eigenvalues and eigenstates of $\rho$ as usual. Note that $d$ is the dimension of the density matrix, and depends on the representation of the density operator we use. 

The two main quantities we need are then the density matrix $\rho$ and its derivatives $\partial_k \rho$. The density operator is defined by Eq. \ref{eq:rho_final}, and we can write the $|\psi_n\rangle$ in the spatial basis as $|\psi_n\rangle = \sum_q \psi_n(u_q, v_q) |q\rangle$, with $\psi_n(u_q, v_q) = \langle q|\psi_n\rangle$. We use a discretized spatial basis defined by the arrays we use to represent the fields in Python. In this case, $q$ indexes the pixel, and $(u_q, v_q)$ are the horizontal and vertical spatial coordinates of pixel $q$. Then, $\psi_n(u_q, v_q)$ is the electric field of the light from source $n$. Note these pixels are distinct from the measurement pixels that we index with $p$, and are defined by our representation of the fields in Python, not the physical measurement. Rather than consider the field in the image plane, we use the field at the aperture of the telescope. This is valid as unitary transformations do not change the QFI, and propagation through a lossless imaging system is unitary, meaning we could calculate the QFI from the fields at any point in their propagation through the telescope \cite{QFImultiparam, reck1994experimental}. Then, 

\begin{equation}
\psi_n(u, v) = A(u, v)e^{-j(k_xu + k_yv)}, \quad k_x=k\frac{x_n}{f}, \quad k_y=k\frac{y_n}{f},
\label{eq:aper_fields}
\end{equation}

where $k=\frac{2\pi}{\lambda}$ is the wavenumber and $A(u, v)$ is the aperture function of the telescope normalized so that $\sum_p |A(u_p, v_p)|^2 = 1$. We assume $A(u, v)$ is a disk with uniform transmission. We note the parameters $x_n, y_n$ appear in this expression and the $b_n$ appear in the density operator definition in Eq. \ref{eq:rho_final}. Together, Eq. \ref{eq:aper_fields} and Eq. \ref{eq:rho_final} allow us to construct the density matrix. 

The derivatives of the density operator with respect to our parameters are 

\begin{equation}
\begin{split}
\frac{\partial\rho}{\partial x_n} &= b_n \frac{\partial|\psi_n\rangle}{\partial x_n}  \langle\psi_n| + b_n|\psi_n\rangle \frac{\partial\langle\psi_n|}{\partial x_n} \\
\frac{\partial\rho}{\partial y_n} &= b_n \frac{\partial|\psi_n\rangle}{\partial y_n}  \langle\psi_n| + b_n|\psi_n\rangle \frac{\partial\langle\psi_n|}{\partial y_n} \\
\frac{\partial\rho}{\partial b_n} &= |\psi_n\rangle\langle\psi_n|
\end{split}
\label{eq:rho_derivs}
\end{equation}

Thus, we can write the density operator and its derivatives using the $|\psi_n\rangle$ and the $\frac{\partial|\psi_n\rangle}{\partial x_n}, \frac{\partial|\psi_n\rangle}{\partial y_n}$. We use autograd in PyTorch to calculate these derivatives in the spatial basis discussed above, but they could easily be calculated analytically using Eq. \ref{eq:aper_fields}.

We could now write the density matrix and its derivatives in the spatial basis, calculate the eigenvectors and eigenvalues, and calculate the QFI. However, the spatial basis dimension equals the number of pixels in our numerical arrays and is quite large to diagonalize numerically. We note that the set of vectors $\{|\psi_n\rangle, \frac{\partial|\psi_n\rangle}{\partial x_n}, \frac{\partial|\psi_n\rangle}{\partial y_n}\}_{n=1}^{N}$ is a set of $3N$ linearly independent vectors and thus spans a space of dimension $3N$. This subspace contains all the vectors we need to represent the density operator and its derivatives. We construct an orthogonal basis for this subspace using a singular value decomposition. Specifically, we define a $3N \times N_{pixel}$ matrix $\boldsymbol{M}$ with each row being one of the $3N$ linearly independent spatial vectors. If the singular value decomposition is  $\boldsymbol{M} = \boldsymbol{U\Sigma V}^{\dagger}$, the first $3N$ rows of $\boldsymbol{V}^{\dagger}$ contain an orthonormal basis for our subspace. We will label these states as $\{|u_i\rangle\}_{i=1}^{3N}$. We can then represent $|\psi_n\rangle, \frac{\partial|\psi_n\rangle}{\partial x_n}, \frac{\partial|\psi_n\rangle}{\partial y_n}$ in this basis. For example, $|\psi_n\rangle = \sum_{i=1}^{3N} \langle u_i|\psi_n\rangle |u_i\rangle$. Thus, we represent all the relevant states as $3N$ element vectors and the density matrix and its derivatives as $3N\times3N$ matrices.

The remaining steps are straightforward. We calculate the eigenvalues and eigenstates of $\rho$, and then use Eq. \ref{eq:SLD} to calculate the SLD operators for each parameter. Then we use Eq. \ref{eq:QFIM_SLD} to calculate the quantum Fisher information matrix, which we invert to obtain the quantum Cram\'er-Rao bound. The singular value decomposition, eigenvalue calculation, eigenvector calculation, and matrix inversion are all done numerically using PyTorch functions.

\section*{Appendix B: Natural Gradient Descent Algorithm}

We use a maximum likelihood estimator as given in Eq. \ref{eq:MLE} to achieve the CRB limit for direct imaging. The formula in Eq. \ref{eq:MLE} can be easily derived given the assumption of independent Poisson noise at each pixel, which results in the measurement distribution

\begin{equation}
p(\mathbf{I}; \boldsymbol{\theta}) = \prod_{p} \frac{\mu_p(\boldsymbol{\theta})^{I_p}e^{-\mu_p(\boldsymbol{\theta})}}{I_p!}.
\label{eq:PoissProb}
\end{equation}
where $\mu_p(\boldsymbol{\theta})$ is the mean photon number at each pixel $p$ and the product is over all pixels. 

In this paper, we implement the maximum likelihood estimate for direct imaging by solving this optimization problem using a natural gradient descent algorithm. This is an iterative optimization approach, where the estimate at iteration $i+1$ is calculated from the previous estimate $\widehat{\boldsymbol{\theta}}^{i}$ as

\begin{equation}
\widehat{\boldsymbol{\theta}}^{i+1} = \widehat{\boldsymbol{\theta}}^{i} - \gamma \boldsymbol{J}^{-1}(\widehat{\boldsymbol{\theta}}^{i}) \nabla\mathcal{L}(\widehat{\boldsymbol{\theta}}^{i}) , 
\label{eq:NG}
\end{equation}

where $\gamma$ is a learning rate and $\mathcal{L(\boldsymbol{\theta})}$ is the negative log likelihood function, which follows directly from the probability in equation \ref{eq:PoissProb} \cite{NGref}. Note this is identical to the standard gradient descent update except that the gradient is rescaled by the inverse of the Fisher information matrix. We find that this rescaling results in better convergence for low Fisher information conditions, such as high contrast scenes, where the likelihood function with respect to some parameters is very flat. Note that this natural gradient descent update is nearly identical to the Fisher scoring algorithm commonly used in estimation theory, except in Fisher scoring the learning parameter is typically equal to 1. In this paper, we use a learning rate of 0.1 for all results. To avoid instability at points where the Fisher information matrix becomes ill conditioned, we regularize the Fisher information matrix before inversion by adding $\epsilon_{reg}\boldsymbol{I}$ to it, where  $\epsilon_{reg}$ is a small number and $\boldsymbol{I}$ is the identity matrix. We use a value of 0.1 for $\epsilon_{reg}$.

The gradients of the loss function can be calculated using autograd in Pytorch, which is convenient and may be necessary depending on the complexity of the measurement model $\mu(\boldsymbol{\theta})$. For the direct imaging case, we derive analytical expressions for the derivatives and use these. The Fisher information matrix is calculated as described in the supplementary material. 

We note that the brightness parameters $b_n$ are strictly positive. It is convenient to enforce this condition in the gradient descent optimization by redefining the parameter vector $\boldsymbol{\theta}$ to contain the log brightnesses, $\ln(b_n)$. These brightnesses are then converted back to the $b_n$ after the optimization, noting that $b_n = e^{\ln{b_n}}$ is positive for any value of the log brightness. 

The algorithm is initialized with equal brightness point sources at random positions inside a square with side length of twice the Rayleigh limit. To avoid local minima, we initialize the algorithms multiple times with different positions to increase the chance of finding the global minima. The results of Fig. \ref{fig:Fig3} use 3 random initializations per trial, and the results of Fig. \ref{fig:Fig4} use 5 random initializations per trial. After the algorithm is run for each initialization, the estimate with the largest likelihood is chosen as the final estimate. The max number of iterations for the algorithm for each initialization is set to 2000, but we terminate the algorithm if the parameter values change by less than a given amount between iterations.

\begin{backmatter}
\bmsection{Funding}
Air Force Office of Scientific Research (FA9550-24-1-0345);
This material is based upon work supported by  
Ministry of Trade, Industry and Energy (RS-2024-00419426). This material is based upon work supported by the Air Force Office of Scientific Research under award number FA9550-23-F-0014 in the amount of \$140,000.

% \bmsection{Acknowledgment} The section title should not follow the numbering scheme of the body of the paper. Additional information crediting individuals who contributed to the work being reported, clarifying who received funding from a particular source, or other information that does not fit the criteria for the funding block may also be included; for example, ``K. Flockhart thanks the National Science Foundation for help identifying collaborators for this work.''

\bmsection{Disclosures} The authors declare no conflicts of interest.

% \bmsection{Disclosures} Disclosures should be listed in a separate section at the end of the manuscript. List the Disclosures codes identified on the \href{https://opg.optica.org/submit/review/conflicts-interest-policy.cfm}{Conflict of Interest policy page}. If there are no disclosures, then list ``The authors declare no conflicts of interest.''

\bmsection{Data availability} Data underlying the results presented in this paper are available from the authors upon reasonable request.

\bmsection{Supplemental document}
See Supplement 1 for supporting content.
\end{backmatter}

% Bibliography
\bibliography{sample}



\section*{Supplementary material}

# Definition of Optical Super-resolution for Telescope Imaging: supplemental document

**Contents**



## 1. POISSON STATISTICS AND DIRECT IMAGING MODEL

In this section, we give an expanded discussion of the Poisson noise assumption specified in the main text and give explicit expressions for the direct imaging measurement defined by $\mu^{DI}$.

To calculate the CRB, we need the probability distribution of the measurements. Our primary consideration to calculate this distribution is the photon number distribution of the collected light. For telescope imaging, light typically comes from thermal light sources. While thermal light is known to exhibit photon bunching, or correlations in photon arrivals, the mean photon number per mode of the incoming light for most thermal sources is much less than one and multi-photon events are rare, leading to independent photon arrivals. This results in the photon number distribution at each detector pixel following an independent Poisson distribution, commonly referred to as shot noise [1, 2]. We note that detector noise also affects the probability distribution of the measurements, but optical detectors are now approaching shot noise limited performance and detector noise is not a fundamental limitation.

The measurement probability given parameters $\boldsymbol{\theta}$ is then

$$p(\mathbf{I};\boldsymbol{\theta}) = \prod_p \frac{\mu_p(\boldsymbol{\theta})^{I_p} e^{-\mu_p(\boldsymbol{\theta})}}{I_p!}. \tag{S1}$$

where $\mu_p(\boldsymbol{\theta})$ is the mean photon number at each pixel $p$ and the product is over all pixels.

We also need an expression for $\mu_p(\boldsymbol{\theta})$. This is determined by the optical measurement system that collects light and directs it to the detector. For a DI measurement using an aberration free imaging system with a circular aperture we have

$$\mu^{DI}(x,y;\boldsymbol{\theta}) = \sum_{n=1}^{N} b_n \mathrm{PSF}(x - x_n, y - y_n)\Delta x^2 \tag{S2}$$

$$\mathrm{PSF}(x,y) = \frac{\pi D^2}{4\lambda^2 f^2}\left[\frac{J_1(D\pi r/\lambda f)}{D\pi r/\lambda f}\right]^2. \tag{S3}$$

We obtain $\mu_p^{DI}(\boldsymbol{\theta})$ by evaluating Eq. S2 at the coordinates of each detector pixel $p$. $(x,y)$ are the horizontal and vertical coordinates at the focal plane of the imaging system and for simplicity we define the target coordinates $(x_n, y_n)$ as the position of the sources on the detector. $b_n$ is defined as the mean photon number from source $N$ detected per integration time. The $\Delta x^2$ factor is the area of the detector pixels and is included so that the units are photons per detector pixel. In Eq. S3, $f$ is the telescope focal length, $J_1$ is the order one Bessel function of the first kind, and $r = \sqrt{(x^2+y^2)}$ is the radial coordinate in the telescope focal plane. The PSF has the shape of an Airy disk and is normalized to integrate to 1 photon.

## 2. CLASSICAL FISHER INFORMATION CALCULATION

As presented in the main text, the formula for the classical Fisher information matrix with Poisson noise is

$$J_{lk}(\boldsymbol{\theta}) = \sum_p \frac{\partial \mu_p(\boldsymbol{\theta})}{\partial \theta_l}\frac{\partial \mu_p(\boldsymbol{\theta})}{\partial \theta_k}\frac{1}{\mu_p(\boldsymbol{\theta})} \quad . \tag{S4}$$

This can be derived from the probability distribution of the measurements in Eq. S1 and the definition of the Fisher information.

Once the measurement model $\mu_p(\boldsymbol{\theta})$ is defined, we need to calculate the partial derivatives of $\mu_p(\boldsymbol{\theta})$ with respect to the parameters. This can be done for almost any measurement model using autograd features in PyTorch. For the direct imaging case, we use the measurement model in Eq. S2 and calculate the derivatives analytically. It is then straightforward to calculate the Fisher information matrix and invert it numerically to obtain the CRB.

## 3. QUANTUM FISHER INFORMATION DENSITY MATRIX DERIVATION

To calculate the quantum Fisher information (QFI), we need an expression for the density operator $\rho$ representing the quantum state of the light entering the telescope. Stars emit thermal light, and the density matrix for thermal light can be written as

$$\rho = \sum_{n=1}^{\infty} \pi_n \rho_n \quad , \tag{S5}$$

where each $\rho_n$ is the density operator for the multimode state with $n$ total photons and $\pi_n$ is the probability of the corresponding $n$ photon state [1, 3]. One form of the QFI $\boldsymbol{\mathcal{F}}(\rho)$ is

$$\mathcal{F}_{lk}(\rho) = \sum_{i,j=0,E_i+E_j\neq 0}^{d-1} \frac{2\mathrm{Re}(\langle e_i|\partial_l \rho|e_j\rangle\langle e_i|\partial_k \rho|e_j\rangle)}{E_i + E_j}, \tag{S6}$$

where the $E_i$, $|e_i\rangle$ are the eigenvalues and corresponding eigenstates of $\rho$, $d$ is the dimension of the density matrix, and $\partial_k$ is shorthand for the partial derivative with respect to $\theta_k$ [4]. Note here we write the QFI as a function of the density operator which itself is a function of the parameters $\boldsymbol{\theta}$. As noted by Tsang [1], since the $\rho_n$ are operators in orthogonal subspaces (subspaces corresponding to photon number $n$), we can write the QFI as a sum of the QFI from each photon number, $\boldsymbol{\mathcal{F}}(\rho) = \sum_{n=0}^{\infty} \boldsymbol{\mathcal{F}}(\pi_n \rho_n)$. Using this fact and Eq. S6, we can obtain the following form of the QFI,

$$\mathcal{F}_{lk}(\rho) = \sum_{n=0}^{\infty} \left[ \pi_n \mathcal{F}_{lk}(\rho_n) + \frac{\partial_l \pi_n \partial_k \pi_n}{\pi_n} + \partial_l \pi_n \sum_{i=0,E_{ni}\neq 0}^{d-1} \langle e_{ni}|\partial_k \rho|e_{nj}\rangle + \partial_k \pi_n \sum_{i=0,E_{ni}\neq 0}^{d-1} \langle e_{ni}|\partial_l \rho|e_{nj}\rangle \right], \tag{S7}$$

where $E_{ni}$, $|e_{ni}\rangle$ are the eigenvalues and eigenstates of $\rho_n$. In Tsang's original derivation of the QFI for point source imaging [4], the probabilities for each photon number were not considered to be parameters (i.e. $\partial_k \pi_n = 0$ for any $k, n$), so only the first term, the QFI for each $\rho_n$ weighted by the $n$ photon probability, was considered. Note that the mean total photon number is related to the $\pi_n$, and inclusion of the total photon number as a parameter results in non-zero derivatives $\partial_k \pi_n$. In this paper, the total photon number is implicitly taken as a parameter as we treat the brightnesses of all the point sources as unknown parameters, so we retain all of the terms in Eq. S7. Numerous generalizations of Tsang's QFI calculation have been considered [5–7], but the authors are not aware of any previous work that has considered all point source brightnesses as parameters to be estimated.

As we note in the section 1, the mean photon number per quantum mode for thermal light is generally very small, typically less than $10^{-2}$ [2]. Tsang argues that the $\pi_n$ for photon number 2 or greater go as the mean photon number squared, and thus drops all the multi-photon states. This assumption results in density matrix

$$\rho = (1-\epsilon)\rho_0 + \epsilon \rho_1. \tag{S8}$$

Tsang further ignores the zero photon state since the derivative of $\rho_0 = |0\rangle\langle 0|$ with respect to any parameter is zero, leaving only the single photon state for the QFI calculation. This approach is standard for the QFI calculation for point source imaging [5–7]. Here, we will retain the 0 photon density operator term, since $\epsilon$ depends on our parameters. We drop the multi-photon states without further extending Tsang's reasoning to our case, and simply note that this zero or one photon model with a small $\epsilon$ parameter results in independent Poisson photon statistics at each detector pixel [1]. This is the generally assumed photon statistics model for telescope imaging unless special care is taken to detect multi-photon events, such as with an intensity

interferometer. For cases where there is significant deviation from Poisson statistics, a different QFI calculation may be needed.

Using Eq. S7 with the density operator in Eq. S8, we obtain

$$\mathcal{F}_{lk}(\rho) = \frac{\partial_l \epsilon \partial_k \epsilon}{1-\epsilon} + \frac{\partial_l \epsilon \partial_k \epsilon}{\epsilon} + \epsilon \mathcal{F}(\rho_1) + \partial_l \epsilon \sum_{i=0, E_{1i}\neq 0}^{d-1} \langle e_{1i}|\partial_k \rho|e_{1j}\rangle + \partial_k \epsilon \sum_{i=0, E_{1i}\neq 0}^{d-1} \langle e_{1i}|\partial_l \rho|e_{1j}\rangle. \quad \text{(S9)}$$

Note that all the terms with derivatives of $\rho_0$ are zero, so only one term from the zero photon state remains. We can combine the first two terms to get $\frac{\partial_l \epsilon \partial_k \epsilon}{\epsilon(1-\epsilon)} \approx \frac{\partial_l \epsilon \partial_k \epsilon}{\epsilon}$ since we assume $\epsilon$ to be close to zero.

In any practical case, we will measure many copies of the quantum state to get enough photons to be able to estimate the source parameters. We are also forced to measure many copies simply because most cameras do not have exposure times shorter than a microsecond, and the coherence time of visible light even with only 1 nm bandwidth is around a picosecond [2]. The total QFI is the sum of the QFIs from each state [4], so if we measure $M$ copies, our QFI expression becomes

$$\begin{aligned} M\mathcal{F}_{lk}(\rho) &= \frac{\partial_l (M\epsilon) \partial_k (M\epsilon)}{M\epsilon} + M\epsilon \mathcal{F}(\rho_1) + \partial_l (M\epsilon) \sum_{i=0, E_{1i}\neq 0}^{d-1} \langle e_{1i}|\partial_k \rho|e_{1j}\rangle + \partial_k (M\epsilon) \sum_{i=0, E_{1i}\neq 0}^{d-1} \langle e_{1i}|\partial_l \rho|e_{1j}\rangle \\ &= \mathcal{F}_{lk}(N_{phot}\rho_1) \quad , \end{aligned} \quad \text{(S10)}$$

where $N_{phot} = M\epsilon$ is the total mean photon number. Thus, we can simply calculate the QFI of the single photon density operator scaled by the total photon number. Note that this scaled density operator is not itself a density operator since it has a trace equal to $N_{phot}$, and the trace can actually change for different parameter values. This differs from previous QFI calculations, which may estimate some brightness parameters, but always leave one brightness parameter fixed by the requirement of unit trace [5–7]. We also note that this removes the explicit dependence of the total QFI on the number of state copies $M$ and the one photon probability $\epsilon$, and depends on just the mean total photon number, which is generally more convenient.

Finally, the single photon density operator is given by $\rho_1 = \sum_{n=1}^{N} p_n |\psi_n\rangle\langle\psi_n|$, where $p_n$ is the probability of the photon coming from source $n$, $N$ is the number of sources, and $|\psi_n\rangle$ is the quantum state of a photon from source $n$ [6]. We now only consider the one photon number state, so the $|\psi_n\rangle$ just describe the spatial probability distribution of the single photon. We use the $n$ index now for source number since we no longer consider photon number.

Our final QFI is $\boldsymbol{\mathcal{F}}(\rho_{eff})$ where

$$\rho_{eff} = \sum_{n=1}^{N} b_n |\psi_n\rangle\langle\psi_n|, \quad \text{(S11)}$$

where $b_n = N_{phot} p_n$ is the mean photon number from source $n$ in the measurement interval. We again note $\rho_{eff}$ is not a true density operator, but this distinction is not important for any of the following calculations.

In section 4, we verify that this expression for the QFI is consistent with Poisson photon statistics.

## 4. QUANTUM FISHER INFORMATION CONSISTENCY WITH POISSON NOISE

In this section, we verify that the QFI calculation given in sections 3 and Appendix A in the main text is correct and consistent with Poisson photon statistics. To do so, we consider the single parameter estimation case, where only one parameter is unknown and all other parameters are considered to be known. In this case, the optimal measurement strategy for estimating the parameter $\theta_k$ is given by the eigenmodes of the corresponding SLD operator $\mathcal{L}_k$ [4]. This optimal measurement is guaranteed to achieve the QCRB in the single parameter case. Note that the optimal measurement is thus different for each parameter, and in general will also be different for different parameter values (i.e. the optimal measurement depends on the scene being imaged).

Let the eigenmodes of the SLD operator $\mathcal{L}_k$ be labeled as $\{|\phi_i^k\rangle\}_{i=1}^{3N}$. Here we consider $3N$ eigenmodes since we represent each SLD operator as a $3N \times 3N$ matrix, but note that the rank of the SLD operators is less than $3N$, so some eigenvalues will be zero, and there will be a space of

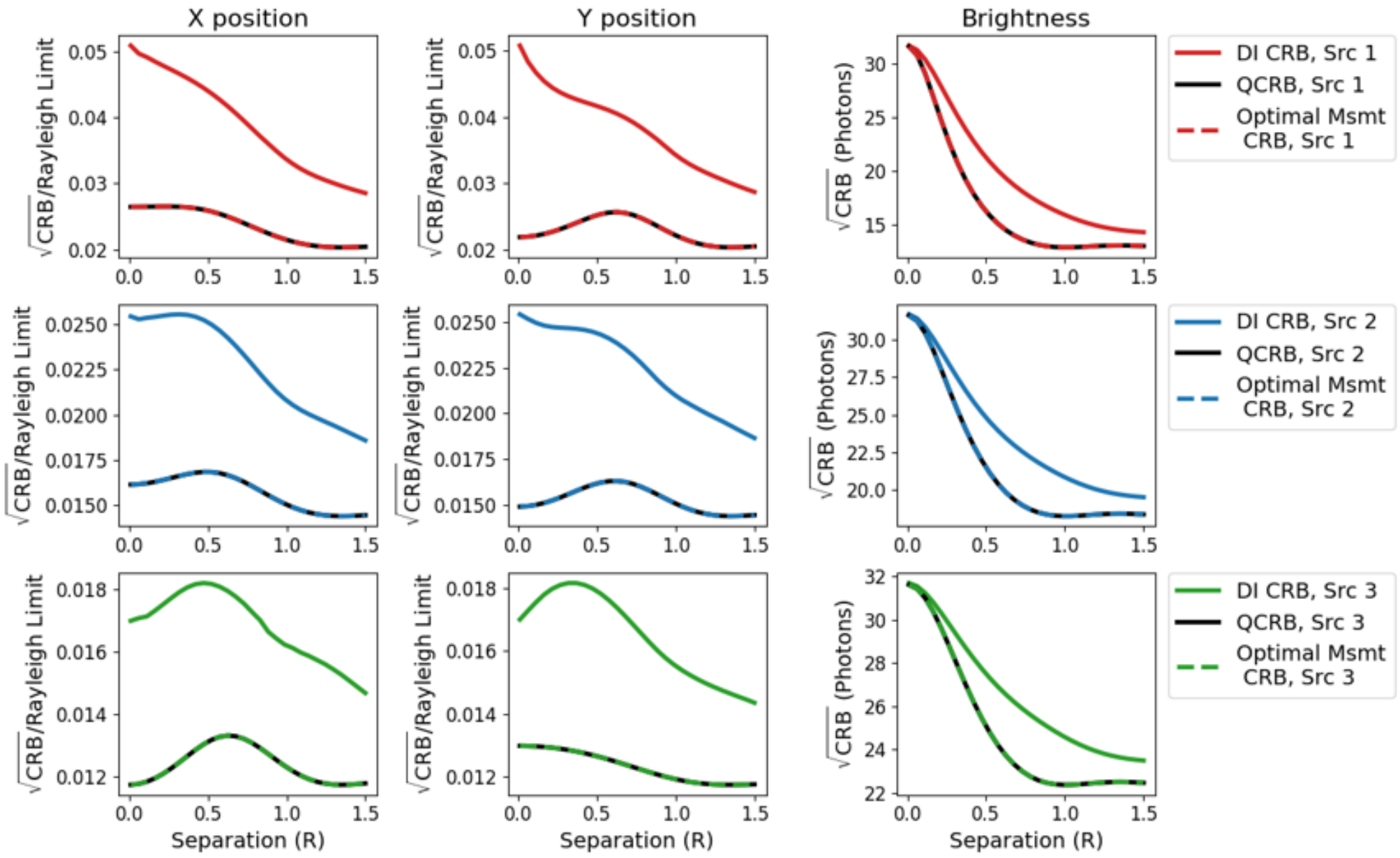


**Fig. S1.** Comparison of the single parameter QCRB to the single parameter optimal measurement CRB and single parameter DI CRB. The optimal measurement is dictated by the eigenstates of the SLD operator for each parameter. Note that the optimal measurement is thus different for each parameter, and is only valid for the single parameter case. We note the QCRB and the CRB for the optimal measurement match, indicating the QCRB is correct and consistent with Poisson photon statistics, as the classical CRB is calculated assuming Poisson noise. For these plots, we use a three source scene with the sources arranged in an equilateral triangle with relative brightnesses [1, 2, 3] and 1000 total photons.

non-unique eigenmodes with eigenvalue zero. The optimal measurement is a projection onto each of the eigenmodes, which corresponds to the measurement model

$$\mu_p^{Proj}(\boldsymbol{\theta}) = \sum_{n=1}^{N} b_n |\langle \psi_n | \phi_p^k \rangle|^2 = \sum_{n=1}^{N} b_n \left| \sum_q \psi_n^*(u_q, v_q) \phi_p^k(u_q, v_q) \right|^2 \tag{S12}$$

where after the second equality we have evaluated the inner products in the discretized spatial basis as described above. For a single source, each pixel of the measurement has a number of photons that is proportional to the modulus squared of the projection of the input field, $|\psi_1\rangle$, onto the basis field of interest $|\phi_p^k\rangle$. For multiple sources, we simply add the photon numbers from each individual source. This is because the fields from each source are incoherent, so we first take the modulus square and then sum over the different sources as opposed to summing the field projections directly. Thus, we can calculate the SLD operators as in Appendix A, calculate their eigenvectors numerically in Pytorch, and then use the measurement model in Eq. S12 to calculate the classical Fisher information as in section 2. Since we assume only a single parameter is to be estimated, the Fisher information reduces to a scalar instead of a matrix, and the CRB is just the reciprocal of the Fisher information. Note that the classical Fisher information directly uses the assumption that the photon number at each pixel is Poisson distributed.

In Fig. S1, we compare the single parameter CRBs for the optimal measurements to the single parameter QCRB for each parameter for a three source scene. The CRB for the optimal measurements match the QCRB in all cases, indicating the QFI calculation is correct and consistent with Poisson noise at the detector.

## 5. POINT SOURCE RMSE CALCULATION

In calculating the error of the estimated position and brightness parameters, we must use some method to match the estimated point sources to the ground truth point sources. For example, if we label the ground truth point sources as "Source 1" and "Source 2", we must decide which of the estimated sources corresponds to "Source 1" and which corresponds to "Source 2". This is done by assigning a cost to matching estimated point source $j$ to ground truth point source $i$

$$C(i,j) = \sqrt{(x_i - \hat{x}_j)^2 + (y_i - \hat{y}_j)^2} + \frac{\beta}{2} |b_i - \hat{b}_j| \tag{S13}$$

where $\beta$ is a heuristic factor that accounts for the different units of the position error (position units) and the brightness error (photon number). We set $\beta$ as the ratio of the mean distance error to the mean brightness error, where the errors are averaged over all the point sources and calculated using an arbitrary assignment of the estimated sources to the true sources. We then pick the final source assignment that minimizes the total cost $C = \sum_j C(i(j), j)$, where $i(j)$ represents the assignment of the estimated source $j$ to one of the ground truth sources $i$.



\end{document}